\documentclass[fleqn,usenatbib]{mnras}

\usepackage{newtxtext,newtxmath}

\usepackage{graphicx}	
\usepackage{subcaption}
\usepackage[T1]{fontenc}
\usepackage{ae,aecompl}
\usepackage{tabularx}

\usepackage{amsmath}	
\usepackage{amssymb}	
\usepackage{natbib, booktabs, multirow, tabularx} 
\usepackage{hhline}

\usepackage{threeparttable}

\usepackage{float}
\restylefloat{figure}

\title[FRB beaming]{Beaming as an explanation of the repetition/width relation in FRBs}

\author[Connor et al.]{%
L.~Connor$^{1, 2}$,
M.~C.~Miller$^{3}$,
D.~W.~Gardenier$^{2, 1}$ 
\\
~\\
$^{1}$ Anton Pannekoek Institute, University of Amsterdam, Postbus 94249, 1090 GE Amsterdam, The Netherlands\\ 
$^{2}$ ASTRON, the Netherlands Institute for Radio Astronomy, Oude Hoogeveensedijk 4,7991 PD Dwingeloo, The Netherlands\\ 
$^{3}$Department of Astronomy and Joint Space-Science Institute, University of Maryland, College Park, MD 20742-2421
}

\date{Accepted XX YY ZZZZ}

\pubyear{2020}

\begin{document}
\label{firstpage}
\pagerange{\pageref{firstpage}--\pageref{lastpage}}
\maketitle

\begin{abstract}
    It is currently not known if repeating fast radio bursts (FRBs) are fundamentally different 
    from those that have not been seen to repeat. One striking difference between 
    repeaters and apparent non-repeaters in the CHIME sample is that the once-off events 
    are typically shorter in duration than sources that have been detected two or more times. 
    We offer a simple explanation for 
    this discrepancy based on a selection effect due to beamed emission, 
    in which highly-beamed FRBs are less easily
    observed to repeat, but are abundant enough to  detect often as once-off events. The explanation 
    predicts that there is a continuous distribution 
    of burst duration---not a static bimodal one---with a 
    correlation between repetition rate and width. Pulse width and opening angle 
    may be related by relativistic effects in shocks, 
    where short-duration bursts have small solid angles due to a large common Lorentz factor. Alternatively, the relationship 
    could be a geometric effect where narrow beams sweep past the observer more quickly, as with pulsars. Our model has implications 
    for the FRB emission mechanism and energy scale, volumetric event
    rates, and the application of FRBs to cosmology. 
\end{abstract}

\section{Introduction}

Fast radio bursts (FRBs) are short-duration ($\mu$s-ms) extragalactic radio transients 
whose origins remain a mystery \citep{cordes2019, petroff_review}. 
To date, approximately 800 FRBs have been detected, 
of which $\sim$\,120 have been published \citep{petrofffrbcat} 
and $\sim$\,700 will be published by the Canadian Hydrogen Intensity Mapping 
Experiment (CHIME) in a forthcoming catalog \citep{fonseca-2020-apj}.
The majority of these have not been seen to repeat. In this paper 
we interchangeably refer to such sources as 
`once-off events' and `apparent non-repeaters', as we cannot 
know that they will not repeat in the future.
There are 20 FRBs that have been found to repeat, all but two of which 
were discovered by CHIME
\citep{spitler2016, chime2019r2, chime19-8repeaters, fonseca-2020-apj, Kumar-2019}.
One of the repeaters, 
FRB\,180916.J0158+65, was found by CHIME to exhibit a 16.35-day periodicity in its 
repetition activity \citep{chime-2020-periodicity}. There is now also a claim 
that the first source found to repeat, FRB\,121102, does so with a tentative $\sim$\,160\,day 
periodicity in its activity level \citep{rajwade-2020}. No FRB periodicity has been detected on 
timescales between 10$^{-3}$\,s and 10$^{3}$\,s that might be associated with 
a neutron star rotation period. 

It remains an open and important question as to whether repeating 
FRBs and apparent non-repeaters form two physically distinct classes. There are 
a number of ways a repeating FRB might be observed as a once-off event, 
including low repeat rate, 
clustered repetition \citep{scholz-2016, connor-2016b, oppermann-2018}, 
or an unfavourable 
luminosity function and telescope sensitivity \citep{connor-2018a,caleb-2019, Kumar-2019}. 
For example, CHIME has only once detected 
the repeater FRB\,121102, so in the absence of previous observations that 
source would appear to be a non-repeater \citep{josephy-2019}.

One curious distinction has emerged between repeaters and non-repeaters 
in the pulse width distribution found by CHIME. They have found that 
repeaters typically emit longer-duration pulses, and once-off FRBs are narrower \citep{chime19-8repeaters, fonseca-2020-apj}. This has been shown  
for the first 12 published apparent non-repeaters \citep{chime2019a} 
and 18 repeating FRBs discovered by CHIME. It was also suggested by 
\citet{scholz-2016}, who noticed that the non-repeating FRBs detected 
at Parkes were shorter in duration than FRB\,121102. Robust comparison 
across different surveys is difficult, but the CHIME FRBs were all 
found on the same telescope by the same pipeline, and it is difficult 
to explain the width/repetition relation with an instrumental selection effect.

We provide an explanation for the broad duration of repeating FRBs 
relative to once-off events using beamed emission. 
If FRBs are beamed with a 
wide distribution of opening angles, and there is a positive correlation 
between opening angle and pulse width, then broad FRBs will have a 
higher observed repeat rate even if the intrinsic 
repetition statistics are the same.

In this paper we first describe our model, including a simple Monte 
Carlo simulation demonstrating the proposed selection effect due to 
beaming. We then speculate on the origin of the required relationship 
between beaming and pulse width, followed by the implications 
for FRB emission, rates, energetics, and future searches for 
repeating sources.

\section{Model}
\label{sect-model}
In our picture, most or all FRBs repeat. The distribution of intrinsic 
repetition rates is currently unknown, and our model remains agnostic to its shape. Instead, 
the key distinction between observed repeaters and FRBs that have only been 
detected once is their beaming angle, not their 
intrinsic repeat frequency (which would be zero in the case where non-repeaters are 
a distinct class of cataclysmic FRBs). The CHIME data show that in roughly one year of observing, which corresponds to $\sim$\,60 hours on each source, there are more once-off FRBs than repeaters, 
and the repeaters are longer in duration. To explain these two facts, 
our model requires two main assumptions: 

\begin{enumerate}
    \item There exists a positive correlation 
          between beaming angle and pulse duration
    \item The intrinsic beaming angle distribution of FRBs
          is such that there are more highly-beamed sources than ones with large 
          opening angles, for the region of widths to which we are sensitive
\end{enumerate}

\begin{figure*}[ht]
	\centering
	\includegraphics[scale=0.5]{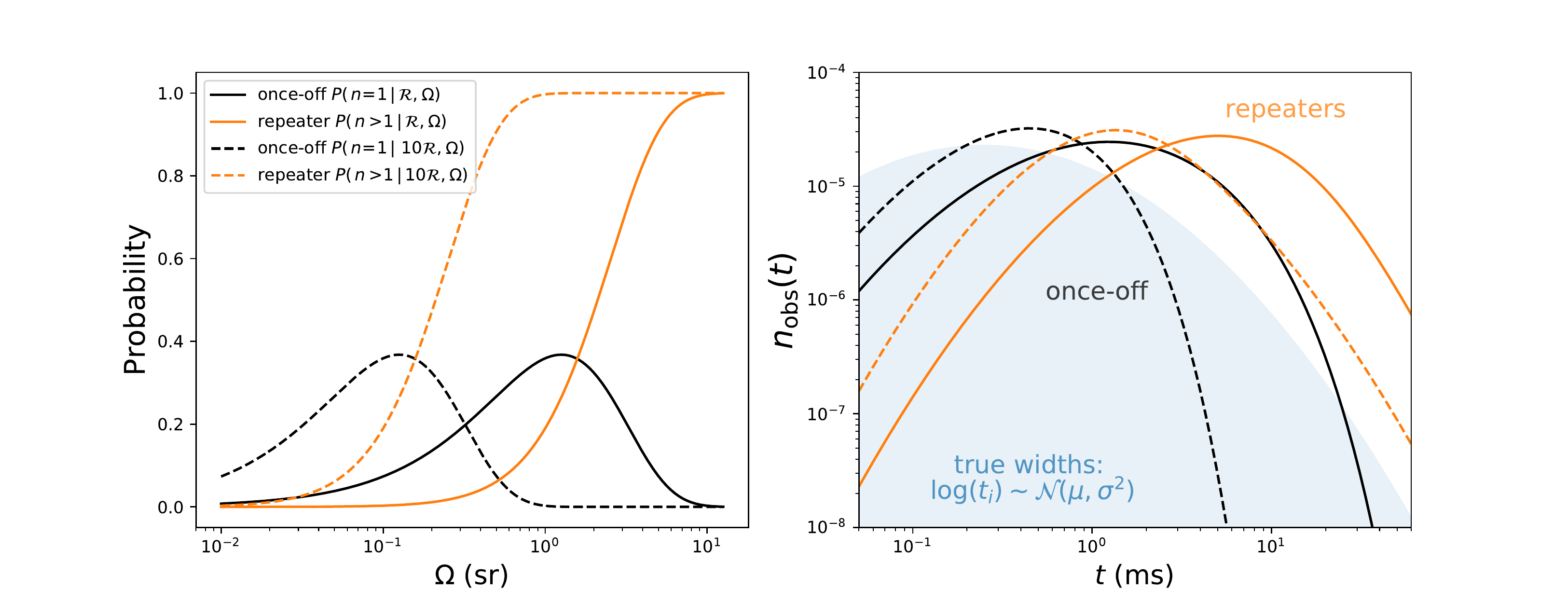}
	\caption{The left panel shows the probability of a beamed 
	FRB emitting in the direction of an observer once (black) 
	and more than once (orange) as a function 
	of opening angle, $\Omega$. We use two intrinsic repeat rates, $\mathcal{R}$ (solid) and $10\,\mathcal{R}$ (dashed), corresponding to an expected value of 5 and 50 bursts per $T_{\rm obs}$, respectively. The right panel 
	shows the observable pulse width distributions of repeaters and apparent non-repeaters, having converted beaming solid angle $\Omega$ to 
	pulse width, $t$, via Eq.~\ref{eq-beaming-time}. 
	There are more broad-duration repeating FRBs, and narrower
	single detections.}
    \label{fig-prob-rep}
\end{figure*}

If sources with beaming solid angle, $\Omega$,
emit repeat bursts in random directions, then on average 
the probability of detecting a burst from a given source is $\frac{\Omega}{4\pi}$. 
In principle, the FRB need not emit uniformly over the full sphere: So long as the directions of its repeat bursts are spread over a solid angle that is larger than $\Omega$, the effect 
still holds. Assuming Poissonian repetition where the $j^{th}$ source has a repetition rate $\mathcal{R}_j$ and opening angle $\Omega_j$, the expected number of bursts in 
observing time $T_{\rm obs}$ is,
\\
\begin{equation}
    N^j_{\rm exp} = \frac{\Omega_j}{4\pi} \mathcal{R}_j T_{\rm obs}.
\end{equation}
\\
\noindent The probability of a source being detected exactly once is
\\
\begin{align}
    P(n\!=\!1\,|\,\Omega_j, \mathcal{R}_j) = 
    e^{-\frac{\Omega_j}{4\pi}\mathcal{R}_j T_{\rm obs}}\frac{\Omega_j}{4\pi}\mathcal{R}_j T_{\rm obs}
    \label{eq-p1}
\end{align}
\\
\noindent and the probability of its repeating twice or more towards the 
observer is,
\\
\begin{align}
    P(n\!\geq\!2\,|\,\Omega_j, \mathcal{R}_j) &= 1 - P(n\!=\!0\,|\, \Omega_j, \mathcal{R}_j) - P(n\!=\!1\,|\, \Omega_j, \mathcal{R}_j)\\
    &= 1 - e^{-\frac{\Omega_j}{4\pi}\mathcal{R}_jT_{\rm obs}}\left (1 + \frac{\Omega_j}{4\pi}\mathcal{R}_jT_{\rm obs}
    \right ).
\end{align}
\\
\noindent To get the beaming angle and pulse width distribution of \textit{detected} events, two more steps are required. 
First, $P(n\!=\!1)$ and $P(n\!\geq\!2)$ must be multiplied by the intrinsic distribution of 
beaming angles, $n_i(\Omega)$. This quantity is the
differential intrinsic beaming angle distribution which 
gives the number of bursts with opening 
angle in each bin $d\Omega$ and is defined as,
\\
\begin{equation}
    n_i(\Omega) \equiv \frac{dn_i}{d\Omega}.
\end{equation}
\\
\noindent Next, we must ask which of those events will actually be detected 
by a radio telescope, after deleterious smearing effects due to finite 
time and frequency sampling. In Fig.~\ref{fig-prob-rep} we plot  
in the left panel the probability that a source is pointed towards the observer once (black) and more than 
once (orange), for two different repetition rates, as a function of 
beaming angle. The right panel shows these probability curves multiplied by the intrinsic 
$\Omega$ distribution, assuming a log-normal 
distribution in beaming angle with a
mean of 0.04\,sr, corresponding to a burst duration mean 
of 200\,$\mu$s. 
To get this pulse width distribution we have assumed a simple mapping between burst duration, $t$, 
and $\Omega$ such that,

\begin{equation}
    \label{eq-beaming-time}
    t = 5\,\mathrm{ms} \times \frac{\Omega}{1\,\,\mathrm{sr}},
\end{equation}
\\
\noindent We speculate on the origin of such a relationship in Sect.~\ref{sect-width-omega}.
Note that the right panel of Fig.~\ref{fig-prob-rep} corresponds to the observed width 
distribution for repeaters and apparent non-repeaters in the absence of instrumental 
smearing and propagation effects such as scattering or plasma lensing. The orange 
curves illustrate that beaming angle provides a selection effect such that 
repeating FRBs will have statistically larger widths than 
once-off events. Conversely, a wide event that has only been detected once 
is more likely to repeat in the future than a narrow apparent non-repeater.

\subsection{Monte Carlo simulation}
\label{sect-MC}
In order to estimate a realistic pulse width distribution of 
detected FRBs, 
we must include non-linear instrumental effects such as  temporal smearing. 
To do this, we have built a simple Monte Carlo simulation. This also 
enables us to add jitter to the pulse width/beaming relationship and 
test different input distributions for $\Omega$, $\mathcal{R}$, 
brightness, and DM. Though we have assumed for Fig.~\ref{fig-prob-rep} 
that the underlying repeat 
rate need not vary between sources, in reality there will be an intrinsic  
distribution $n_i(\mathcal{R})$ which may include some true non-repeaters,
i.e. weight at $\mathcal{R}=0$.

\begin{figure*}
	\centering
	\includegraphics[scale=0.5]{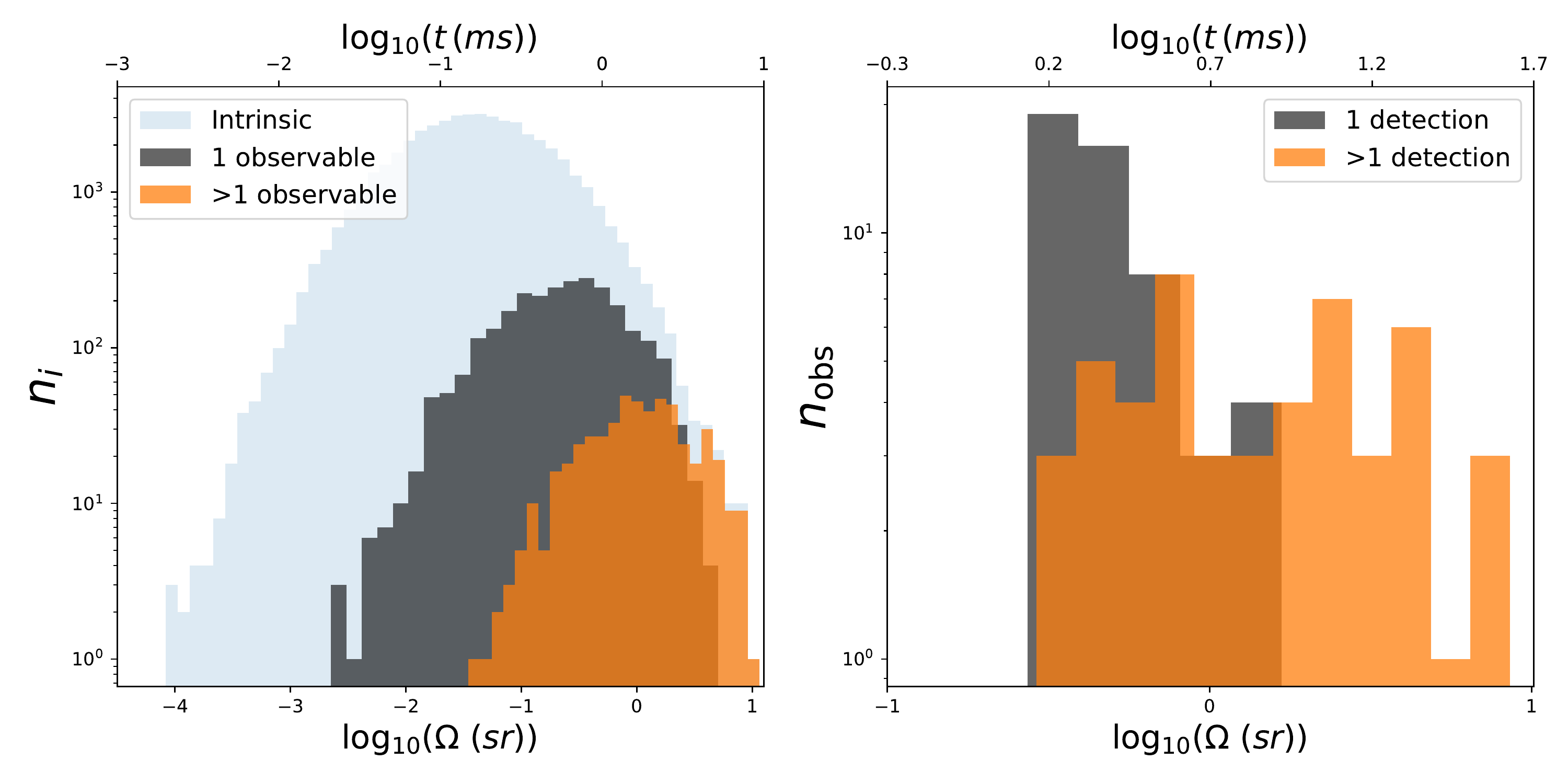}
	\caption{The blue histogram in the left panel shows the beaming angles of 100,000 simulated 
	        FRBs, all with the same intrinsic repeat rate. The black and orange counts 
	        are FRBs that were `observable' once and more than once, respectively, thanks to 
	        their favourable viewing angle and their being within the temporal observing window. The right panel 
	        shows the detected distributions of pulse widths, after accounting for 
	        instrumental smearing with a 
	        CHIME-like back-end. The detected repeaters are statistically significantly wider than the 
	        more-numerous single-detection FRBs in this realization.}
	 \label{fig-MC}
\end{figure*}

We start by 
simulating 100,000 FRBs, all with the same intrinsic repeat rate but with a 
broad distribution of beaming angles, as shown in Fig.~\ref{fig-MC}. 
We then simulate 1000 repeat bursts for each source,
using Poissonian statistics and with an emission direction that is drawn 
randomly from a uniform distribution on the sphere. Each FRB 
is observed for 20--30 hours, drawn from a uniform distribution.
We then check which if any of their repeat
bursts would be detectable with CHIME.

If a burst from a given FRB was emitted within its $T_{\rm obs}$ and with favorable 
pointing, 
it was deemed `observable' (left panel of Fig.~\ref{fig-MC}), which is to say the observer line-of-sight fell within that FRB's top-hat beam during the pre-defined time window.
Within that subset of observable bursts, we take the linear relationship 
between $\Omega$ and $t$ used in Eq.~\ref{eq-beaming-time}. 
We then apply instrumental smearing effects, assuming a 
CHIME-like back-end. We assume for simplicity that all FRBs have DM=1000\,pc\,cm$^{-3}$. A pulse with duration $t_i$ as it arrives at our telescopes will be smeared to 

\begin{equation}
    t_{\rm obs} = \sqrt{t^2_i + t^2_s + t^2_{\rm DM}},
\end{equation}

\noindent due to the finite time and frequency sampling of radio telescopes. Here, $t_s$ is the instrument's sampling time and 
$t_{\rm DM}$ is the timescale associated with 
intra-channel dispersion smearing, given by, 

\begin{equation}
    t_{\rm DM} = 8.3\times10^{-3}\,\mathrm{DM}\,\,\frac{\Delta\nu_{\rm MHz}}{\nu^3_{\rm GHz}}\,\,\,\,\,\mathrm{ms}.
\end{equation}

\noindent We use the values for CHIME's back-end because they have the 
largest sample of repeaters, and that is where the repetition/width 
relation is most pronounced. Its 
current sampling time is $t_s=0.983$\,ms and its frequency channel width 
in MHz is $\Delta\nu_{\rm MHz}=0.024$. Using DM=1000\,pc\,cm$^{-3}$ 
and a central frequency in GHz of $\nu_{\rm GHz}=0.6$, the minimum 
detection width for an FRB in our simulation is 
$t_{\rm obs}=\sqrt{t^2_s + t^2_{\rm DM}}\approx1.35$\,ms. That is why the 
right panel of Fig.~\ref{fig-MC} has no detections below that value, despite 
the large number of sub-millisecond simulated events. 

Instrumental smearing also decreases the pulse's signal-to-noise ratio (S/N) 
and lowers its chances of being detected. To account for this in our simulation, we apply 
the corresponding reduction in S/N. A pulse whose brightness corresponds to some S/N, $s_i$,
in the absence of smearing, will be detected with a S/N,

\begin{equation}
    \label{eq-snr}
    s_{\rm obs} = s_i \left ( \frac{t^2_i}{t^2_i + t^2_s + t^2_{\rm DM}} \right )^{1/2}.
\end{equation}
\\
\noindent We assume for now that there is no correlation between width and 
brightness, and draw each value $s_i$ from a Euclidean distribution in brightness
with $n(s_i)\propto s^{-5/2}_i$. In Sect.~\ref{sect-width-omega} we discuss 
how this assumption may not hold if the relationship between $\Omega$ and $t$ 
is due to relativistic beaming, which would cause narrow bursts beamed in our 
direction to be brighter. 
If the pulse's resulting $s_{\rm obs}$, as computed in Eq.~\ref{eq-snr}, is above a S/N threshold,  $s_{\rm min}$, we 
label it a `detection'. In the right panel of Fig.~\ref{fig-MC} we show the
subset of simulated bursts that were observable and still above $s_{\rm min}$ after smearing. We have ignored a 
potential selection effect by using a single DM value. 
If once-off FRBs are in fact brighter, then they will be visible at 
greater distances and may have higher DMs, which can lead to 
detection biases. But based on the current 
CHIME sample of repeaters and 
single-detections, whatever differences might emerge between 
their DMs will be relatively small and their distributions 
will overlap significantly.

It is clear that 
detected repeaters are wider in duration than the apparent non-repeaters. The resulting distributions are similar to the observed widths in CHIME, in that apparent non-repeaters 
are more abundant and cluster around the smearing width
and are narrower in duration.

For the sake of isolating the beaming selection effect proposed in this paper,
we have used a simplified model and must include the following caveats. 
FRB repetition is often clustered and not described by a homogeneous Poisson 
process \citep{scholz-2016, oppermann-2018, gourdji2019}. This increases the 
variance on the number of detected bursts in an observing window, 
even if the mean remains the same, $N_{\rm exp} = \frac{\Omega_j}{4\pi} \mathcal{R}_j T_{\rm obs}$. We have assumed a delta function 
distribution in DM with $n_i(\rm DM) = \delta\!\left (\mathrm{DM} - 1000 \right )$ when 
that is known to not be the case. We have also used a perfect mapping between $\Omega$ and pulse 
duration $t_i$. Still, we have experimented with adding noise 
to the $\Omega(t_i)$ function such that there is jitter in the mapping, 
and find that so long as there is a correlation between beaming angle 
and duration, there is a difference between the detectable 
widths of repeaters vs. apparent non-repeaters. 
To tie this to observations, we have also tried adding noise to the 
$t_i$/$\Omega$ relation for an individual repeating FRB, not just 
from source to source. We 
have taken the fractional root-mean-square error (RMSE) on the pulse widths of  
FRB\,180916.J0158+65 as the spread 
for our simulation, because it is the CHIME repeater with the 
greatest number of detections. 
For FRB\,180916.J0158+65 this value is 
$\sim$\,0.6. With this level of noise, we find that repeaters
are noticeably broader than single-detections, even from just the 
first 19 repeating sources. It is possible that our estimated 
RMSE is biased low, because very narrow and very broad FRBs are 
more likely to be missed.
We also tested different 
distributions in $\Omega$ and DM and recover the same effect as long as 
the two main assumptions in Sect.~\ref{sect-model} are met. Finally, 
the pulse widths reported by CHIME are fitted widths rather than the 
maximum-S/N boxcar that was used to discover the FRB \citep{chime2019a, chime19-8repeaters, fonseca-2020-apj}. They are able to 
do this effectively because of their large fractional bandwidth, which can 
disentangle intrachannel dispersion smearing, scattering, 
and intrinsic duration, which have different frequency dependencies.
This is why some of CHIME's published FRB widths are shorter 
than the sampling time of the 
instrument. We have opted instead to use the final smearing width, because 
our simulation did not produce dynamic spectra of individual bursts 
to fit, which is why the red points in Fig.~\ref{fig-chime-repeaters} 
go to shorter timescales than the orange points. Nonetheless, the width/repetition effect is expected with 
both methods.

\section{Discussion}

\subsection{Predictions}

In our model, repeating FRBs and those that have only been detected 
once are not two fundamentally different source classes. Therefore, 
we do not expect a bimodal distribution of FRB widths in which 
repeaters have a characteristic duration and non-repeaters have a different 
one. This 
is in contrast to GRBs, which can be divided into two classes along 
burst duration, where short-hard bursts are typically less than 
2\,s and long-soft bursts are longer than 2\,s  
(\citeauthor{Kouveliotou-1993}, \citeyear{Kouveliotou-1993}; see the review in \citeauthor{berger-2014}, \citeyear{berger-2014}). 

Unlike with 
GRBs, our model suggests that frequent 
repeaters are simply the 
long-duration tail of the non-repeater distribution. The repeater width distribution will move towards shorter durations 
as the exposure to each source increases, because that leads to a 
greater probability of seeing a previously once-off source repeat. 
If, instead, repeaters and non-repeaters 
have distinct and \textit{static} width distributions, 
there are several summary statistics that could be used. 
For example, the bimodality coefficient, $\beta$, is defined between 
0 and 1, where $\beta=5/9$ corresponds to a uniform distribution and
greater values can imply multimodality. It 
can be used as a test statistic, assuming the 
underlying distribution is generated by a mixture of two normal
distributions\footnote{https://en.wikipedia.org/wiki/Multimodal\_distribution} (whether in $t$ or $\log t$). 
It is computed as,

\begin{equation}
\beta = \frac{g^2 + 1}{k + \frac{3(N-1)^2}{(N-2)(N-3)}}
\end{equation}

\noindent where $N$ is the number of samples in the dataset, 
$k$ is excess kurtosis, and $g$ is the sample skewness. Another method 
called Kernel Mean Matching (KMM) has been used in astronomical 
datasets such as globular cluster metallicity and 
GRB duration \citep{ashman-1994}. In the case of FRBs, the 
observed width distribution can depend strongly 
on the time/frequency resolution of the 
survey back-end \citep{connor-2019}. Before a test for 
bimodality can be done, burst-duration selection effects 
must be understood and a large-enough sample of 
temporally resolved FRBs must be obtained. 
CHIME will be able to 
do this if their selection function can be measured, 
because they can save raw voltage data and can fit for pulse widths 
below their instrumental smearing timescale. If the 
presence of two peaks were found, and they do not change 
with increased exposure, this would be good 
evidence against the claim that all FRBs are repeaters 
with a continuum of repetition frequencies.

\begin{figure}
	\centering
	\includegraphics[width=0.4\textwidth]{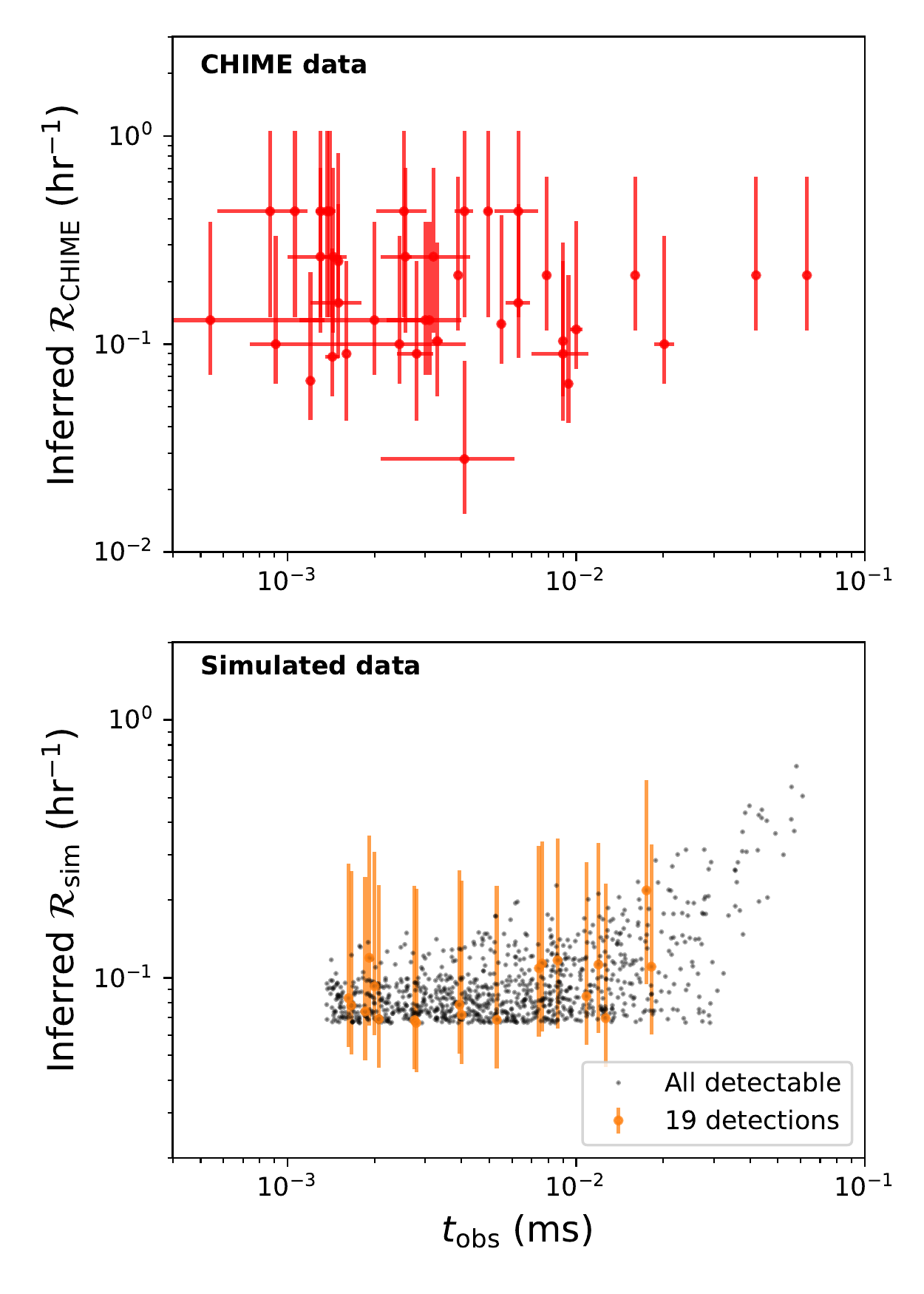}
	\caption{The inferred repetition rate vs. pulse width 
	of CHIME repeaters (top panel) and simulated events 
	(bottom panel). In both cases, the first 19 repeater detections do not 
	result in a significant correlation. The Poissonian error 
	bars we use are lower limits for the true 
	error bars, which are presently unknowable due to the biases and 
	selection effects described in the text.}
	\label{fig-chime-repeaters}
\end{figure}

We also expect a positive correlation between FRB duration and 
repetition rate. The strength of this correlation will depend 
on the mapping between beaming angle and intrinsic pulse width, as well as observational selection 
effects. Again, this is because our population is not 
divided into repeaters and non-repeaters. That is, 
even within the set of detected repeaters, we expect 
those that repeat more frequently to have wider beams.
From our simulation, we find that such a correlation persists 
so long as the $\Omega/t$ relationship is not made too noisy 
and a large-enough collection of repeaters have been observed for 
more than an average repeat period. This is shown in the bottom 
panel of Fig.~\ref{fig-chime-repeaters}, where 
there is a discernible relationship for the large number 
of black points, but none for the first 19 detections
in our simulation.

We looked for such a relation in the CHIME repeater data, 
but found no significant correlation between mean repetition rate and pulse width (top panel of Fig.~\ref{fig-chime-repeaters}). The correlation was tested using a Pearson product-moment correlation, but was not found to be constraining  either in the linear or in a logarithmic space.
At this point, that is unsurprising for the following reasons. 
With just 19 repeaters observed for a median $\sim$\,23 hours, 
many sources have only repeated two or three times. This combined 
with the uncertainty of CHIME's exposure to each source (from unknown 
beam shapes, day-to-day sensitivity, etc.), 
leads to large uncertainties in their repeat rate, 
which we take as $\mathcal{R}^j_{\rm CHIME}=N_j/T^j_{\rm obs}$. Furthermore, 
the inferred repeat rate of sources that have been detected 
just a few times is highly biased. Suppose there were 1000 FRBs
all with the same intrinsic Poissonian repeat rate, and they were 
observed for a duration less than their common repeat period. 
The first dozen repeater detections would necessarily have 
a much higher inferred repeat rate than their 
underlying repetition frequency. Therefore, we should expect 
the CHIME repeaters that have been detected fewer than 5 times
to repeat less frequently in the future 
than they have thus far. Finally, 
as discussed in Sect.~\ref{sect-model}, temporally clustered, 
non-Poissonian repetition increases the error bars on 
the number of bursts detected in an observing window, furthering uncertainty in repeat rate.
The Poissonian error bars we use for the 
red points in Fig.~\ref{fig-chime-repeaters} are therefore 
lower limits on the true uncertainty.

In Fig.~\ref{fig-chime-repeaters} the 
orange points show the first 19 FRBs detected in our Monte Carlo simulation. 
As with CHIME repeaters, many have not been detected more than a 
couple of times and there is no strong correlation with pulse width. 
In order to establish our proposed correlation, more repeating 
sources are needed and each source must be observed for longer to decrease 
uncertainty on their repeat period. The black points 
in the bottom panel of Fig.~\ref{fig-chime-repeaters} give 
an idea of the broader trend and the number of repeaters required to measure this correlation. In our simulation 
(and similarly with CHIME), the repeaters have only been 
observed for 20--30\,hours, so there is a floor in the repeat rate at 2 per 30\,hours. That floor can be seen in the plot's black points. 
Until those sources are observed for longer, they do not offer 
much information on the repeat rate/width correlation, 
but the broader, more repetitive sources do.
A better estimate of their repeat rates could be obtained by 
tracking telescopes and with coming years of CHIME exposure.

One consequence of any model in which repeaters and 
once-off FRBs are drawn from the same population, such as ours, is that 
the pulses ought to have similar individual burst structure. 
The first repeating source to be discovered, FRB\,121102, often 
emits FRBs with a characteristic `march down' in their dynamic spectra, such that 
adjacent sub-pulses peak at subsequently lower 
frequencies \citep{hessels2019}. This has also been detected in 
some, but not all, repeating FRBs discovered by CHIME \citep{chime2019r2,chime19-8repeaters, fonseca-2020-apj}. 
We might then expect that with sufficient 
time/frequency resolution and S/N, some narrow apparent non-repeaters
would have similar structure in their dynamic spectra, but on 
shorter timescales. This is possible 
now that many FRB back-ends allow for the preservation 
of raw voltage data that can be coherently dedispersed
\citep{farah2018, chime19-8repeaters, bannister19, ravi-2019-DSA}. 
We emphasize that while this should be true for our model, 
it will also be the case for any scenario in which 
repeaters and apparent non-repeaters come from similar sources in similar environments, not only 
under the beaming selection effect we have 
put forth.

Depending on the physical origin of a beaming/pulse width 
correlation, there will be more model-specific predictions, some of 
which we touch on in the next subsection. Here we have 
described only the most generic consequences of a situation 
in which broadly beamed FRBs are more easily detected as repeaters.


\label{sect-width-omega}
\subsection{Origin of the $\Omega$/$t$ relationship}
Beamed emission occurs in astrophysical sources when collimated 
beams of particles are moving at speeds close 
to $c$. If those particles are moving 
with a Lorentz factor, $\Gamma$, radiation is seen 
by the observer within an angle $\alpha\sim\Gamma^{-1}$. 
If those relativistic particles form a pencil beam, 
so too will the radiation and 

\begin{equation}
    \Omega\sim\alpha^2\propto\Gamma^{-2}.
\end{equation}

\noindent If the particles are confined to a 
thin sheet, the contraction is effectively only 
in one dimension \citep{katz-2017} and we get,

\begin{equation}
    \Omega\sim\alpha\propto\Gamma^{-1}.
\end{equation}

The purpose of this paper is to propose a simple 
model for the origin of the longer durations of 
repeating FRBs relative to apparent non-repeaters rather 
than offer a unique emission mechanism or progenitor. 
Our model requires that FRBs are differentially beamed 
and that their opening angle scales with pulse width.
It also requires that repeat bursts from the same source 
point in different directions over a larger 
area than its own beaming solid angle $\Omega$.
While we do not attempt to explain this phenomenon 
at the emission level with high certainty, below 
we provide examples of how the $\Omega$/$t$ relationship
could emerge in the context of previously-proposed 
FRB models.

\subsubsection{Relativistic temporal modulation}

In a subset of FRB models, relativistic plasma is sporadically expelled 
from magnetars which then collides with 
surrounding material or the magnetar's wind, producing 
radio emission $\sim$\,$10^{11}$--$10^{15}$\,cm away from the star---well 
outside of its magnetosphere \citep{lyubarsky-2014,beloborodov-2017, metzger-2019, margalit18,  beloborodov-2019}. 
This is preferentially expected to occur in 
young, hyper-active systems, which are thought to generate 
magnetic flares more frequently than the older magnetars we observe in
our own Galaxy. Such models provide a natural connection 
between burst duration and observability, due to relativistic effects. 

In these emission models, the ultra-relativistic shock 
has Lorentz factor, $\Gamma_{\rm sh}$. The resulting electromagnetic 
radiation is compressed in time by the Doppler effect and 
undergoes relativistic beaming, in the observer's frame. From \citet{beloborodov-2019}, the burst duration in the observer's frame is,
\\
\begin{equation}
    t_i(r) \approx \frac{R_{\rm GHz}}{c\,\Gamma_{\rm sh}^2},
\end{equation}
\\
\noindent where $R_{\rm GHz}$ is the distance from the 
star at which the observer-frame emission peaks at 
GHz frequencies. The beamed solid angle is,
\\
\begin{equation}
    \Omega \approx \frac{\pi}{\Gamma_{\rm sh}^2},
\end{equation}
\\
\noindent and therefore a natural connection between 
beaming solid angle and pulse width emerges,
\\
\begin{equation}
    t_i \approx \frac{R_{\rm GHz}}{c\,\pi}\,\Omega.
\end{equation}
\\
\noindent In this scenario, both $t_i$ and 
$\Omega$ scale with $\Gamma_{\rm sh}$, which 
may vary from burst to burst and from 
source to source. 

The variation in Lorentz factors between sources 
must be larger than the variation within a source 
between bursts, so that frequently-observed repeating 
FRBs are broader in duration (smaller $\Gamma_{\rm sh}$) on average.
There is already some evidence of this being the case if FRB data 
are interpreted within the shocked gas framework \citep[see Fig. 3]{margalit-2020a}. Within the flaring magnetar model, 
there is an alternative explanation for the 
proclivity of wider FRBs to repeat often that is more
intrinsic to the source. Burst width 
increases with the density of material around the neutron star, 
with which the relativistic flare must collide, and therefore 
width and repeat rate are correlate positively \citep{margalit-2020a}.

\subsubsection{Rotation period \& opening angle}
Another explanation for the connection between 
opening angle and pulse duration comes from models in which 
beamed emission is sweeping past the observer.
It is not known if the beamed radio emission from 
pulsars forms a circularly symmetric pencil beam 
\citep{rankin-1990} or if it is a fan beam that
is narrow in the direction transverse to the observer 
but broad in the orthogonal direction \citep{michel-1987,wang-2014, oswald-2019}. Even in the fan beam model, 
the emission does not span 180\,deg in the latitudinal 
direction, so highly-beamed emission is still more difficult 
to observe.
Empirically, \citet{rankin-1990} found that 
pulsar widths correlate with the neutron star 
rotation period in seconds $P_{\rm sec}$ as,
\\
\begin{equation}
    t_i \approx 6.6\times10^{-3}\,\,\mathrm{s}\,\,\sqrt{P_{\rm sec}}\,\,\,.
\end{equation}
\\
\noindent The duty cycle, $W$, 
which is proportional to the radio 
beam's opening angle transverse 
to the observer, scales as $\frac{1}{\sqrt{P}}$. 
Therefore old, non-recycled pulsars with large 
periods have smaller opening angles. It must be noted, 
however, that there may be selection biases related to 
periodicity searching and pulsar duty cycle. 
Nonetheless, it has been proposed that one reason old pulsars 
become unobservable is not just that they become too 
faint as they approach the death line, but 
that their narrow beams are less likely to 
point in the direction of the observer 
as their magnetic and spin axes become more aligned \citep{johnston-2017}. 

That scenario 
is similar to the selection effect we have
described in this paper except that in the case of pulsars, older sources 
have spun down so much that their pulse durations are wider, despite having 
narrower opening angles. If FRBs are produced by a rotating, pulsar-like object \citep{cordes-2016, kumar-lu-2017}, 
then we require wide opening angles to correspond to long duration bursts, 
so $W$ and $P$ should not anti-correlate in the same way as 
Galactic pulsars. 
Within this framework, our model would also require that each repeat burst 
is emitted sporadically at a range of pulse phases, since 
periodicity has not been seen in repeating FRBs on short timescales ($10^{-3}$--$10^{3}$\,s). One tension with this explanation comes from 
the fact that the polarisation position angle (PA) is known to be flat across 
the pulse \citep{michilli2018, chime19-8repeaters}, even though 
a PA swing is expected in the standard rotating vector model. 
If the emitting region is close to the rotational 
equator and the beaming angle is much less than a radian, 
then the PA may be relatively constant, but the magnetic field geometry may 
offer clues about a possible connection between opening angle and 
duration.

In \citet{katz-2017} it was suggested that while FRBs are often thought 
to radiate nearly isotropically and repeat with low duty cycle, 
it may be that they are almost always emitting, but the emission 
is beamed and only occasionally so in the direction of the observer. 
In this `wandering narrow beam' model, the burst duration 
depends on both opening angle and the angular speed at which 
the beam drifts past the line of sight. As long as the distribution 
of angular speeds does not dominate the observed pulse width, 
such a scenario would lead to frequent repeaters being longer 
duration than once-off FRBs, due to our proposed selection 
effect.

Relativistic beaming is 
also expected in models that invoke orbiting planets 
or asteroids around neutron stars \citep{mottez-zarka-2015, mottez-2020}. 
It was proposed 
that FRBs could be generated in the Alfv\'{e}n wings of orbiting bodies 
interacting with highly-relativistic pulsar winds, 
analogous to the electrodynamics of the Jupiter-IO system. 
In their model the radio flux is concentrated within 
$\Omega\propto\Gamma^{-2}_W$, where $\Gamma_W$ is the 
Lorentz factor of the pulsar wind. The duration of this pulse 
in the observer frame is determined by the angular size of the beamed emission as it sweeps by the observer, so 
$t_i\propto\Gamma_W^{-1}$ \citep{mottez-2020}. 



\subsection{Implications}
FRBs can have peak flux densities as large as the brightest
single pulses from Galactic pulsars, despite coming from 
roughly a million times farther away. The 10--15 
order-of-magnitude gap in pseudo-luminosity can be explained 
by beaming, which is central to our model. But while 
beamed emission alleviates burst energetics, it exacerbates 
the already-high volumetric rates of FRBs \citep{Nicholl-2017, ravi-2019-repeaters}. Therefore, 
FRB emission models that invoke significant beaming 
must offer a way of producing $\frac{4\pi}{\Omega}$ 
times more bursts. We note, however, two useful quantities that are beaming 
invariant: FRB brightness temperature and the total power 
emitted by the source population. The former implies that 
the necessity of coherent emission is not relaxed by beaming, 
and the latter has implications for the global energy requirements
of FRB emitters in the Universe.

In the case of relativistic beaming, it may be possible to 
detect short-duration, high-$\Gamma$ FRBs to greater distances, 
assuming the rest-frame burst luminosity is the same as broader FRBs. Conversely, frequently-repeating FRBs would be closer.
We note that the repeater FRB\,180916.J0158+65 is at $z\approx 0.034$, 
just 150\,Mpc away \citep{marcote20}, 
and has an average pulse width that is wider than 
typical CHIME non-repeaters \citep{chime19-8repeaters}. 
A repeating FRB 
detected within $\sim$\,30\,Mpc would be an ideal system to 
search for electromagnetic radiation beyond the radio and to
precisely study the host environment.
Depending on the ability of DM to predict $z$, 
a large sample of repeating 
FRBs may have a lower average DM than once-off events in 
the CHIME data after accounting for selection effects 
like dispersion smearing. 

If the FRB width distribution continues to show the duration/repetition trend, 
broad once-off FRBs ought to be followed up to search for repeat bursts, independent 
of our explanation for the phenomenon's origin.
CHIME is a transit instrument and cannot point, but northern-hemisphere telescopes 
like Apertif \citep{artsso20} and the the Karl G. Jansky Very 
Large Array (VLA) \citep{law-2018} could 
follow up wide single-detection FRBs found by CHIME, assuming that 
width is not dominated by scattering or instrumental smearing. 
For example, 
FRB\,121102 has only been detected once by CHIME and 
the one detection was 
very broad, at 34\,ms \citep{josephy-2019}. If that source were not already known to be a repeater, FRB\,121102 
would have been an obvious candidate to follow up based only on the width criteria.
As CHIME observed 
between 400 and 800\,MHz, higher-frequency coverage 
could offer interesting insights into frequency dependence 
of pulse width and 
repetition activity.
There appears to be a width/frequency relation in 
FRB\,121102, with a shorter timescale at higher 
frequencies \citep{Gajjar2018}.
That correlation appears to be unrelated to 
instrumental smearing and may 
be a useful probe of the FRB emission 
mechanism and beaming.

The apparent proclivity of once-off FRBs to be 
shorter in duration and more abundant than 
longer-duration, more repetitive sources has implications 
for FRB applications. Many proposed methods of 
using FRBs as probes of cosmology \citep{weltman-2019, Madhavacheril-2019}, the intergalactic medium (IGM) \citep{mcquinn2014, vedantham2019}, or fundamental physics \citep{munoz2016, eichler-2017} 
require larger numbers of sight lines.
This can be hard to attain if the FRBs to which you are 
sensitive are frequent repeaters, because the number 
of distinct sources is inversely proportional to 
repetition rate for a fixed all-sky FRB rate (10$^3$ FRBs 
per sky per day could be produced by just forty sources with 
hourly repetition, for example).
Therefore, if there 
exists a population of narrow FRBs that are currently being 
missed due to instrumental smearing \citep{connor-2019}, surveys that hope to detect a large number of distant sources for 
FRB applications must mitigate these effects in their 
telescope back-end to go after sub-millisecond events. 
Even if the detection rate 
of a given telescope were not increased by changing its 
time/frequency resolution, the number of sight lines 
could increase substantially.

\section*{Acknowledgements}
We thank the essential services members of the Netherlands 
for allowing us to carry out this research from our homes during 
the COVID-19 outbreak.
The research of LC and DWG was supported by 
the European Research Council under the European Union's Seventh Framework Programme
(FP/2007-2013)/ERC Grant Agreement No. 617199 (`ALERT').
MCM thanks the Radboud Excellence Initiative for supporting his stay at Radboud University. We thank our referee for constructive feedback 
that has improved our paper.
We thank Joeri van Leeuwen, Jason Hessels, and Lorenzo Sironi
for helpful conversations. We also 
thank Ben Margalit, Shu-Qing Zhong, and Jonathan Katz for 
helpful comments on the first draft of this manuscript.

\section{Data availability}
All of the analysis presented in this paper ought to 
be reproducible using the text in our manuscript and 
the Python code in our publicly available 
Jupyter notebook\footnote{https://github.com/liamconnor/frb-beaming}.







\bibliography{main}
\bibliographystyle{mnras}





\label{lastpage}
\end{document}